\pdfoutput=1
\documentclass[aps,prb,onecolumn,groupedaddress]{revtex4}

\usepackage{float}
\usepackage{amsmath}
\usepackage{amssymb}
\usepackage{siunitx}
\usepackage{chemformula}
\usepackage{graphicx}
\usepackage{dcolumn}
\usepackage{bm}
\usepackage{siunitx}
\usepackage{lmodern}

\usepackage[T1]{fontenc}
\usepackage[utf8]{inputenc}
\usepackage{textcomp}
\usepackage{epstopdf}
\usepackage{natbib} 
\usepackage{hyperref}
\usepackage{ulem}
\usepackage{soul,color}
\makeatletter
\begin{document}

\title{Radiation damage effects in amorphous zirconolite}
\author{A. Diver, O. Dicks}
\affiliation{School of Physics and Astronomy, Queen Mary University of London School of Physics and Astronomy, Mile End Road, London, E1 4NS, UK}
\author{A. M. Elena, I. T. Todorov}
\affiliation{Daresbury Laboratory STFC UKRI, Scientific Computing Department, Keckwick Lane, Daresbury WA4 4AD, Cheshire, UK}
\author{K. Trachenko}
\affiliation{School of Physics and Astronomy, Queen Mary University of London School of Physics and Astronomy, Mile End Road, London, E1 4NS, UK}

\begin{abstract}
We report the results of a large-scale modelling study of radiation damage effects in the nuclear waste form zirconolite. We particularly focus on the effects of radiation damage in amorphous zirconolite and have developed a new way to analyse the damaged structure in terms of local coordination statistics. On the basis of this analysis, we find that the amorphous structure responds to radiation damage differently from the crystal. Amorphous zirconolite is found to be "softer" than crystalline zirconolite with a much larger number of atoms becoming displaced and changing coordination during a 70 keV cascade. The local coordination and connectivity analysis shows that the amorphous structure continues to evolve as a result of repeated radiation damage, changes which cannot be identified from globally averaged properties such as pair distribution functions. We also find large density inhomogeneities at the local level which we suggest may play an important role for future developments in nuclear waste storage. Finally, we find a correlation between the changes in enthalpy and local coordination, suggesting that measurements of enthalpy change can be linked quantitatively to structural radiation damage. Our results raise an interesting possibility of whether an evolution of the amorphous structure due to radiation damage can converge to a new equilibrium amorphous regime, posing the fundamental question of what that structure may be.

\end{abstract}

\maketitle

\section{Introduction}

Radiation damage effects on the long term material properties of crystalline phases proposed to store high level nuclear waste, including excess Pu, are of much interest to the nuclear energy community\cite{hyatt}. A crucial material of interest is zirconolite (CaZrTi$_2$O$_7$), one of the main three titanate host phases in SYNROC \cite{ringwood_kesson_ware_hibberson_major_1979} alongside hollandite and perovskite. Zirconolite is specifically of interest as compared to the latter two as it is the main phase responsible for the incorporation of actinides, which are highly radioactive and propose the biggest issue with regards to long term safe storage. Due to this there has been much work on quantifying the effects of radiation damage in zirconolite, mainly focused on the transition from the crystalline to the amorphous state. Studies include investigation of natural zirconolite samples, exposed to irradiation over geological timescales of up to 550 million years \cite{lumpkin_ewing,ewing1983alpha}, as well as pristine samples subjected ion beam irradiation experiments. \cite{ewing1992amorphization}. 

In recent years there also has been simulation based work \cite{chappell2012structural,foxhall1,veiller2002molecular}, modelling the structural changes in crystalline samples due to collision cascades, focusing especially on the effects of radiation induced amorphization. However, to our knowledge, there has been no work modelling high energy radiation cascades in already amorphised zirconolite samples. Understanding radiation damage in amorphous zirconolite is essential for improving our understanding of the irradiation hardness and stability of this material. Indeed, a crystalline wasteform will undergo transition to a fully amorphous state after a few thousand years \cite{weber1998radiation} due to alpha irradiation, whereas the storage time for excess Pu and its decay chain is upwards of 1,000,000 years \cite{weber1998radiation}. Therefore, nearly all damage will take place in the waste form's amorphous phase which must remain an effective barrier for a long period of time.

In this work, we use large-scale molecular dynamic simulations of high-energy radiation events in both crystalline and amorphous zirconolite in order to understand the difference in their response to radiation damage. We develop and use a new way to analyse the damaged structure in terms of local coordination statistics. On the basis of this analysis, we find that the amorphous structure responds differently to that of the crystalline one. Amorphous zirconolite is found to be softer than tits crystal phase in terms of recovery of defects, as defined in the amorphous structure. Our local coordination and connectivity analysis shows that the amorphous structure continues to evolve as a result of radiation damage, and shows that this is not visible in globally averaged properties such as pair distribution functions. We also find large density inhomogeneities at the local level which we propose play an important role in future nuclear waste storage solutions as reduced local density channels may serve as fast diffusion pathways for radioactive actinides. Finally, we find a strong correlation between the changes in enthalpy and local coordination, suggesting that measurements of enthalpy change can also be used to infer the extent of structural radiation damage. 

Crucially this study demonstrates that the amorphous structure continues to evolve with increasing radiation damage. Therefore, our results raise an interesting question of whether the detected evolution of the amorphous structure in response to radiation damage may {\it converge} to a new amorphous structure, given enough simulation time. This poses a question of fundamental importance to the physics of the amorphous state \cite{zallenbook}, namely what is the nature of this new amorphous state? The questions posed here should act as the starting point to stimulate future experimental and modelling work. 

\section{Methods}

The molecular dynamics package \texttt{DL\char`_POLY\char`_4}, which uses domain decomposition to simulate systems with large numbers of atoms in parallel \cite{todorov_smith_trachenko_dove_2006}, has been used to simulate the effects of the alpha decay of Pu in crystalline and amorphous zirconolite. 

The focus of this work is to model the recoil of the heavy daughter nucleus from an alpha decay event. The alpha-decay of $^{239}$Pu results in the recoil of uranium atom with approximately 70-100 keV of kinetic energy. The recoil produces most of the structural damage in the system in the form of a collision cascade consisting of thousands of permanently displaced atoms. The accumulation of these cascades leads to amorphization of the crystalline structure. To probe the extent of these atomic displacements and radiation's effect on the structure of the material, we have developed on-the-fly analysis tools and implemented them in the DL\_POLY\_4 package, including time-resolved coordination analysis \cite{diver2020evolution}. All collision cascades were carried out using Archer supercomputing facility. Each cascade, simulated for 180 ps in a cell containing 13,365,000 atoms, was carried out using 1200 MPI processes, taking approximately 20 hours of walltime to complete.

A Buckingham potential which has previously been used in studies of zirconolite \cite{foxhall1} is used to model the inter-atomic forces between atoms. These pairwise potentials are then joined at short-range with a repulsive ZBL \cite{ziegler1985} stopping potential via the use of a switching function $f$ to give the mixed interaction $\phi_{\rm mix}$ as $\phi_{\rm mix}=f\phi_{\rm ZBL}+(1-f)\phi$, where $\phi$ is the Buckingham potential, $f=1-\exp(-(r_m-r)/\xi)/2$ for $r<r_m$ and $f=\exp(-(r_m-r)/\xi)/2$ for $r>r_m$. The values of parameters in the switching function are chosen to give the smoothest fit and are shown in Table \ref{table:switchingpara}.

The use of 13 million atom-sized simulations boxes is used to contain the multiple overlapping cascades investigated in this study. All collision cascades were carried out using the NVE ensemble. We used a variable timestep starting at 1 fs, to consider the initial high velocities of atoms at the beginning of the collision cascades. The starting positions of each recoil are within a few Angstroms of the same coordinate position within the simulation box, where a Zr is replaced with a U atom. Each overlapping cascade occurs along the same initial direction from the same position with the only difference being the initial structure is taken from the end of the previous cascade. This process enabled us to maximize the amount of damage in the same region of the simulation box to probe the effects of damage saturation in this region. 

The amorphous zirconolite structure was produced using the melt-quench technique (similar to previous work \cite{diver2020evolution}). The crystalline system was melted at 6000 K for 100 ps, then cooled to 300 K with a quench rate of 10 Kps$^{-1}$.The system was then relaxed for 100 ps at 300 K using the NPT Ensemble. Radiation damage-induced amorphization is often accompanied by a decrease in density. Here, we observe a 4\% increase in the volume of the amorphous structure after relaxation when compared to the crystal. This is in good agreement with the experimentally observed 4-6 \% volume expansion \cite{clinard1984self,wiss2007helium} of zirconolite as it becomes completely amorphous.

\begin{table}[H]
\centering
\begin{tabular}{ccc}

Atom pair & $r_m$ [\AA]& $\xi$[\AA]\\
\hline

Ca-O& 1.10 & 0.15  \\
Ti-O &1.10  & 0.25\\
Zr-O &1.35  & 0.15\\
O-O &1.60  & 0.15\\
\hline

\end{tabular}
\caption{Switching function parameters used for each individual atom pair}
\label{table:switchingpara}

\end{table}

\section{Results and analysis}

\subsection{Collision cascades and main structural changes}

Quantifying the effect of radiation damage on a material's atomistic structure is challenging, especially if the objective is to link single recoil events with experimentally measurable quantities. A common measure of damage, or introduced disorder to a system, the pair distribution function (PDF), undergoes only negligible changes as a result of a radiation damage event in simulations of both amorphous and crystalline zirconolite systems. This is due to the relatively small number of displaced atoms (approximately 6500) when compared with the entire system of 13 million atoms. 

\begin{figure}
\centering
\includegraphics[width = 8cm]{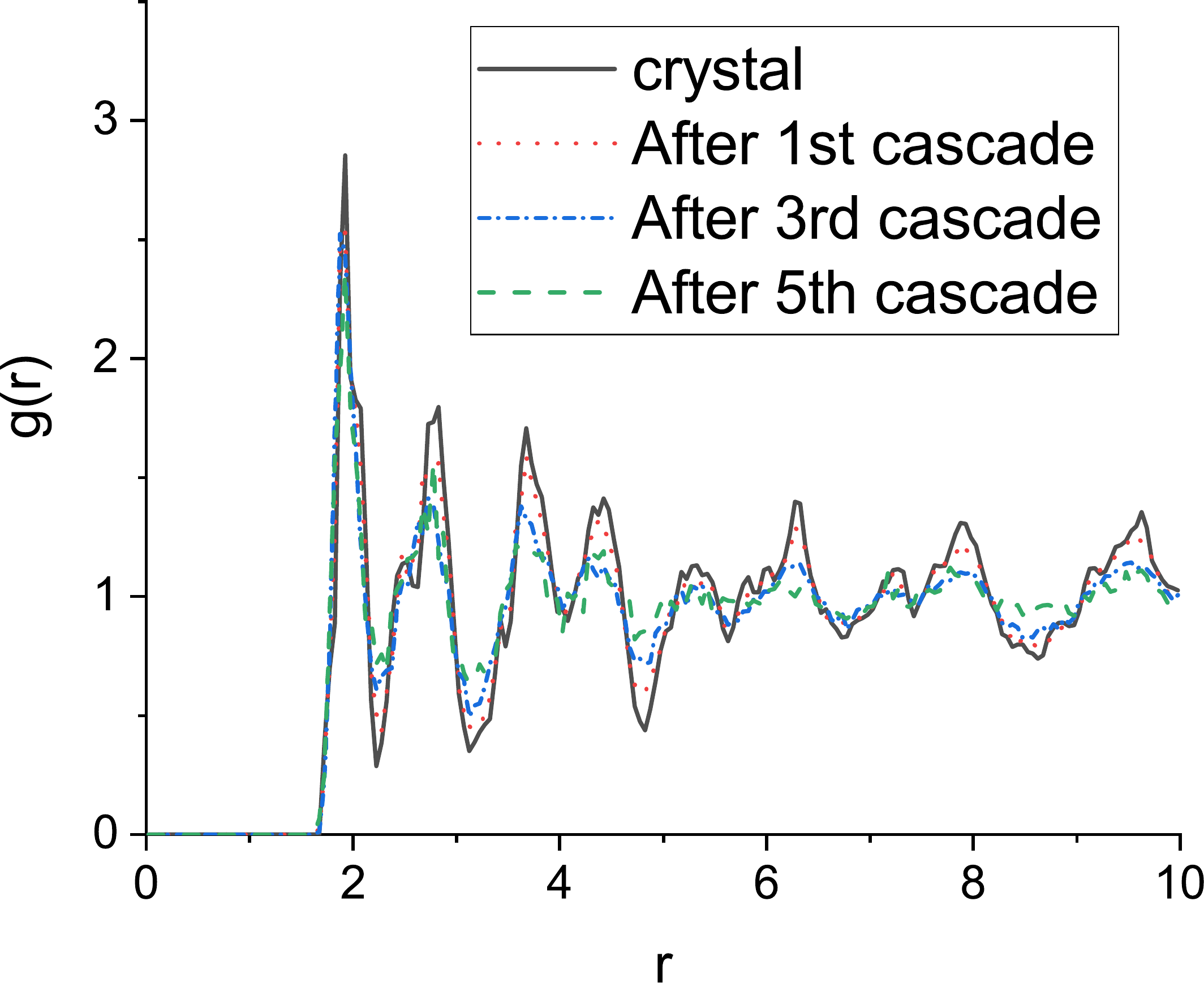}
\caption{Total radial distribution function for crystalline zirconolite in the overlapping region before and after cascade 1, 3 and 5}
\label{fig:crydamregiongr} 
\end{figure}

However, it is possible to zoom in and calculate a PDF of the most damaged region in the cell, centred on the position in the structure where the cascades overlap the most. On this basis a PDF was calculated for both the crystalline and amorphous systems in a small sub-region of the cell, an area containing 6500 atoms. In Fig. \ref{fig:crydamregiongr} the effect of increasing amorphization of the region after multiple cascades can be observed with the loss of long range order in the PDF. However, complete amorphization of the region is not observed, even after 5 cascades. This is in contrast with our earlier study in zircon \cite{diver2020evolution} showing complete amorphization after 4 overlapping cascades. Experimental papers report that the dpa (displacements per atom) to cause amorphization at room temperature in both zircon and zirconolite are roughly the same order of magnitude, 0.2-0.6 dpa for zircon \cite{ewing1995zircon} as compared to 0.67-1.2 dpa for zirconolite \cite{smith1997situ}. The larger dpa required for complete amorphization of zirconolite when compared to zircon is therefore consistent with our modelling results showing zirconolite's greater resistance to amorphization.

In the amorphous system prepared by melt quench, the effect of radiation damage is unnoticeable after 5 cascades except for a slight decrease in the height of the first peak in the PDF, as shown in Fig. 2. The insensitivity of the PDF to radiation damage, even in the small cascade region of amorphous zirconolite, demonstrates the need for other measures to quantify structural damage in amorphous materials.
 
\begin{figure}
\centering
\includegraphics[width = 8cm]{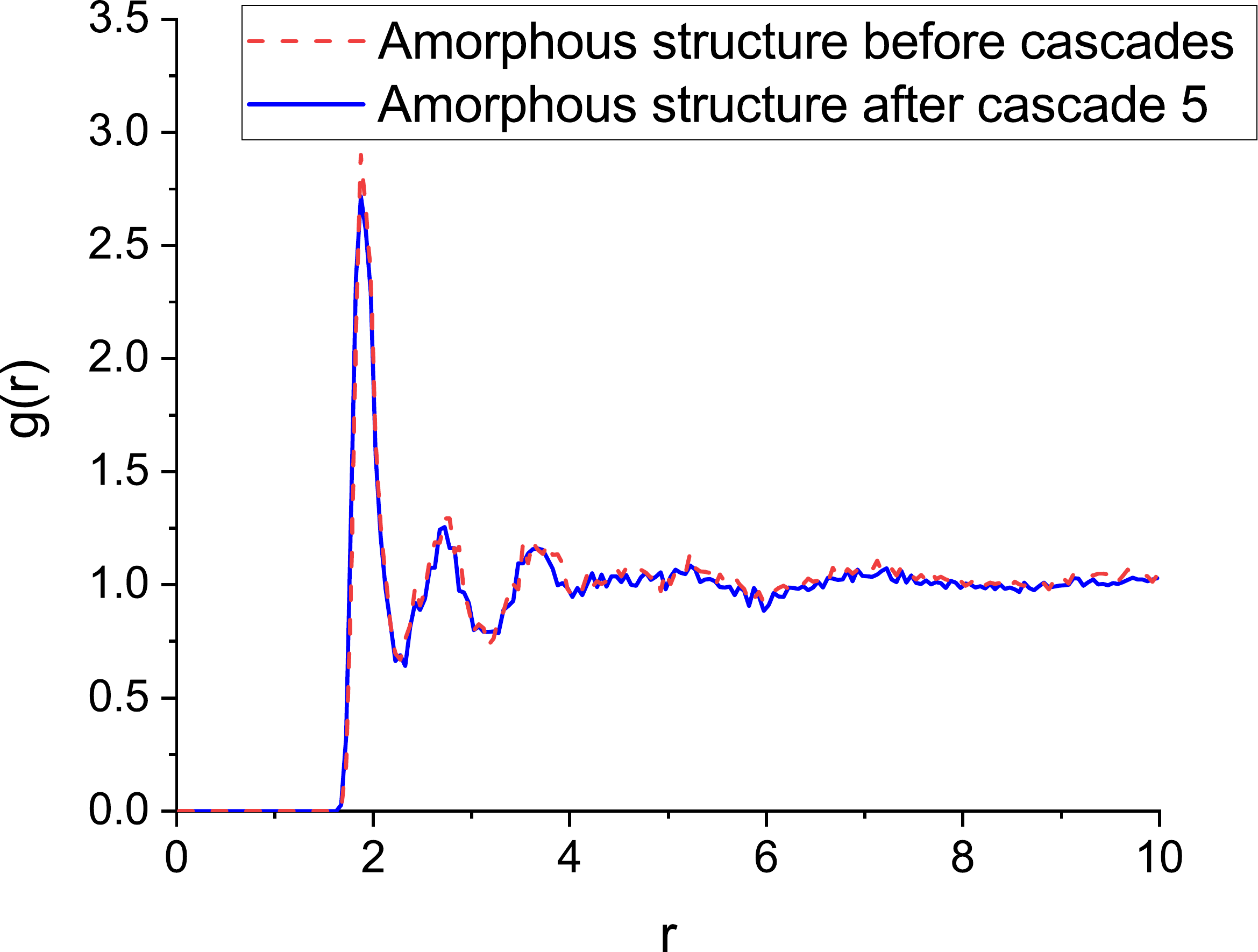}
\caption{Total radial distribution function for amorphous zirconolite in the overlapping region before and after cascade 5}
\label{fig:amrdamregiongr} 
\end{figure}

\subsection{Coordination statistics in crystalline zirconolite due to multiple cascades}

\begin{figure}
\centering
\includegraphics[width = 8cm]{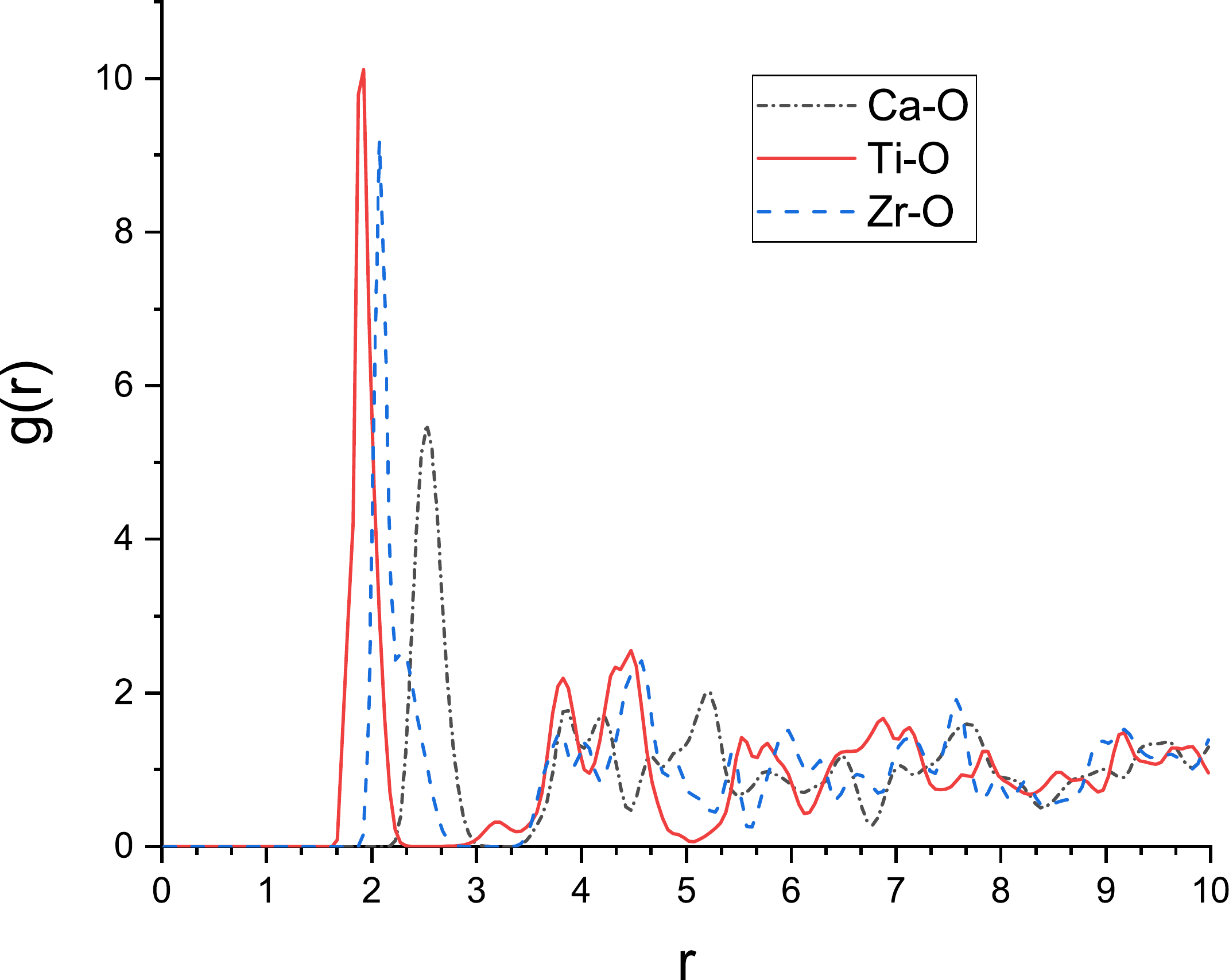}
\caption{Pair distribution functions of Ca-O, Zr-O and Ti-O pairs in crystalline zirconolite before radiation damage}
\label{fig:grcoordpairs} 
\end{figure}

In this section, radiation cascades in crystalline zirconolite are analyzed with the use of coordination based statistics, as in our previous work \cite{diver2020evolution}. In summary, displaced atoms are defined as those that change their local coordination bond network and are translated by more than a chosen cutoff. In zirconolite, defining defects on the basis of local coordination is not straightforward, but the coordination distribution of each atomic species can be nevertheless monitored during the cascades to see how the structure evolves. 

To build the local atomic coordinations between atoms we use cut offs of 2.50 \si{\angstrom}, 3.00 \si{\angstrom} and 3.20 \si{\angstrom} respectively for Ti-O, Zr-O and Ca-O pairs. These values are taken from the first minimum after the first maximum in the relevant pair distribution functions from Fig. \ref{fig:grcoordpairs}. These values are then used in our coordination displacement definition with the specific atom needing to also be displaced by a minimum  of 1.0 \si{\angstrom}, 1.0 \si{\angstrom}, 1.2 \si{\angstrom} and 1.2 \si{\angstrom} respectively for Ti, Zr, Ca and O atoms. The values for the minimum displacement come from the first peak in Fig. \ref{fig:grcoordpairs} and approximately half that value, with the O value chosen to be the same as the largest bond pair distance.

\begin{figure}
\centering
\includegraphics[width = 6cm]{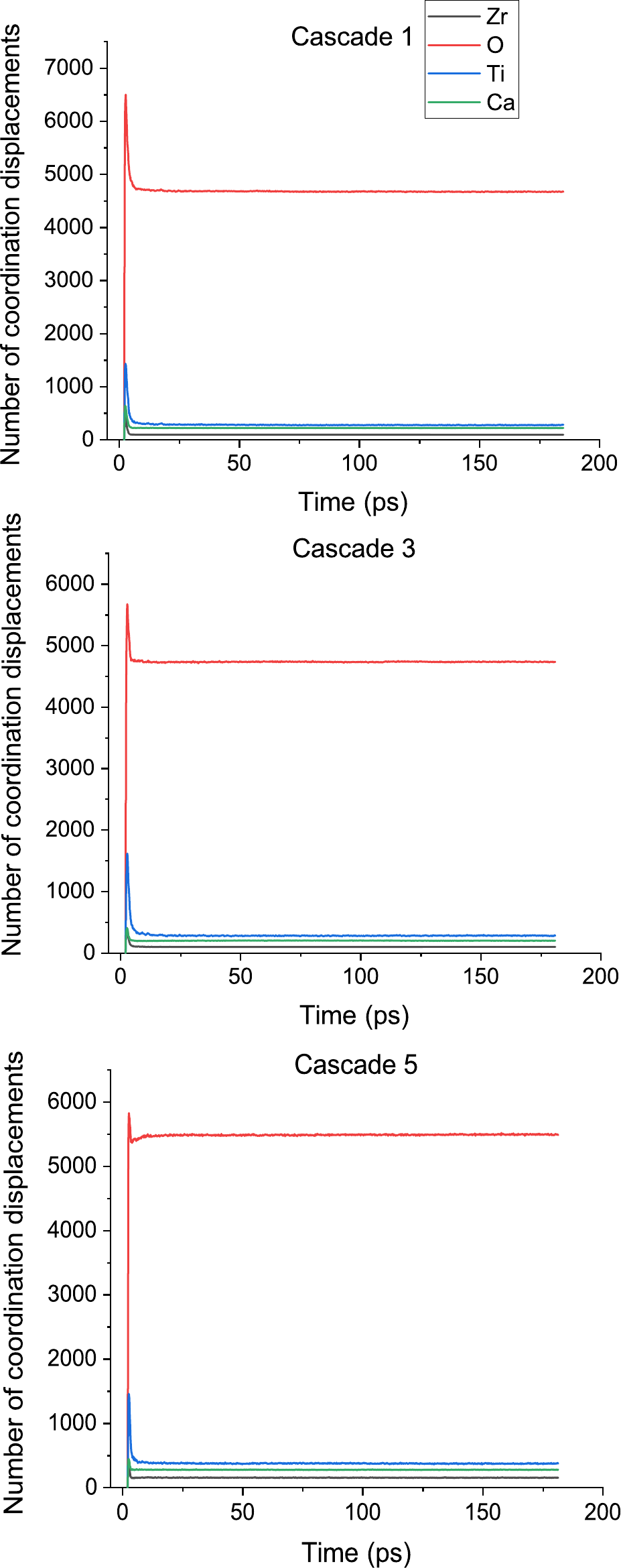}
\caption{Number of coordination displacements for Ti, O, Ca and Zr atoms during collision cascades 1,3 and 5 in crystalline zirconolite}
\label{fig:crydis}
\end{figure}

In Fig. \ref{fig:crydis} we show the time evolution of the coordination displacements caused by cascades 1 (the initial cascade), 3 and 5. The vast majority of the atoms being displaced are oxygens, approximately 5 times more than all the other atomic species combined. This is consistent with previous simulations used to calculate the threshold displacement energies in crystalline zirconolite, \cite{veiller2002molecular} showing that O atoms have the lowest threshold displacement energy (15 eV) when compared to Ca (25 eV), Zr (48 eV) and Ti (45 eV).

As more overlapping cascades are simulated, the total number of displaced atoms per cascade rises, with an increase of around 10\% more coordination displacements after the 5th cascade compared to the first. An increase in the number of defects produced by overlapping cascades in zirconolite has been observed previously \cite{chappell2012structural}, although differences in definition of defects and displacements, and a smaller simulation cell size, make it hard to compare magnitudes. The important finding here is that the amorphous zirconolite structure is more susceptible to damage than the initial crystal phase.

Importantly, the magnitude of O atom coordination displacements in Fig. \ref{fig:crydis} clearly shows the significant decrease in the recovery of the peak at around 1 ps. This suggests as a region becomes more and more damaged it becomes much harder for atoms to return to their crystalline lattice position, promoting amorphization. Small recovery in the number of displaced O atoms after 5 overlapping cascades is consistent with the perseverance of long range order in the local PDFs in Fig. \ref{fig:crydamregiongr}.

\begin{figure}
\centering
\includegraphics[width = 10cm]{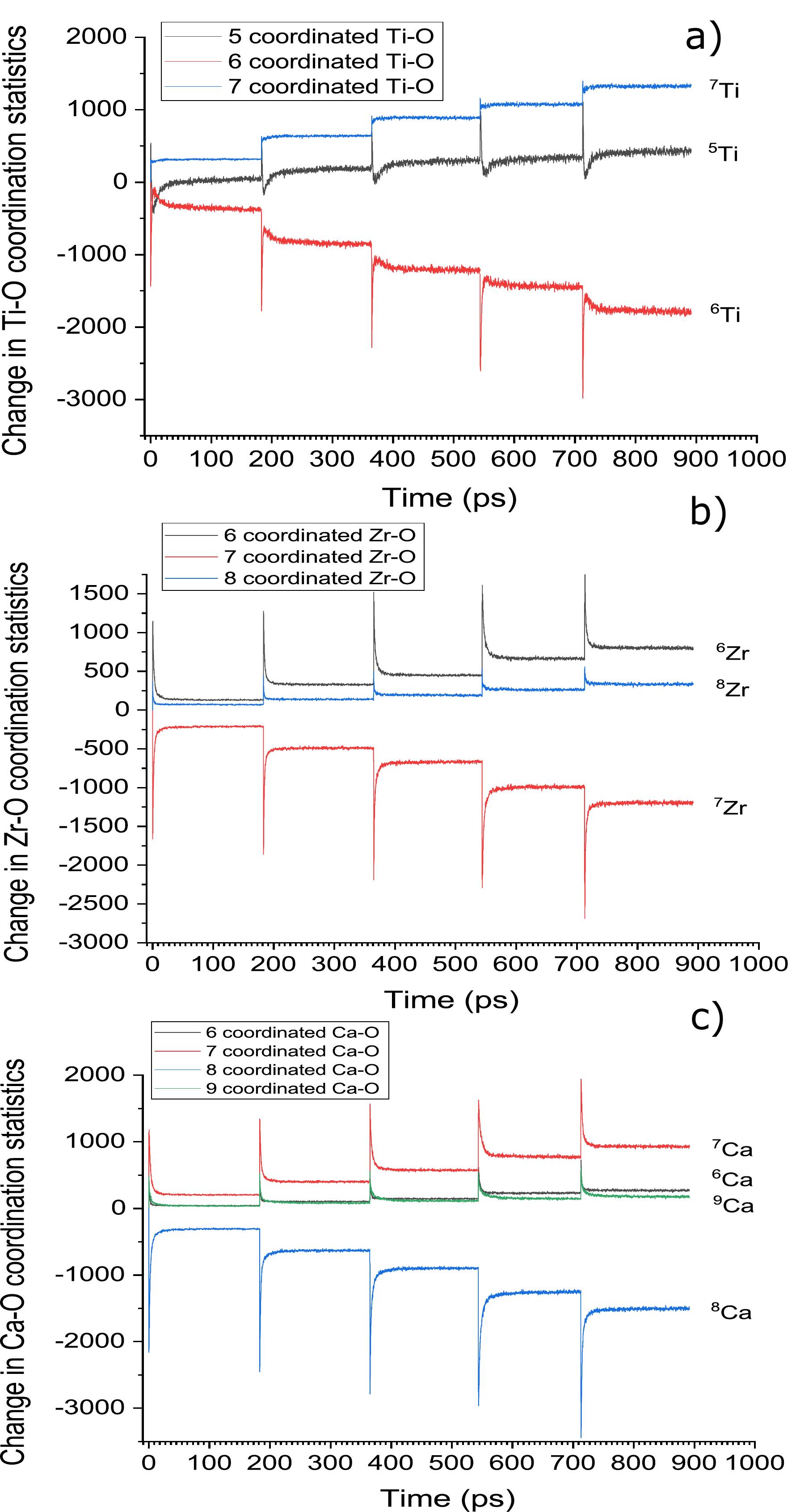}
\caption{Change in atom pair coordination during consecutive overlapping collision cascades in crystalline zirconolite. a) Ti-O , b) Zr-O and c) Ca-O. }
\label{fig:crystats}
\end{figure}

The evolution of the zirconolite crystalline (as well as amorphous) structure can be further analysed from the overall change in coordination statistics of each cation-oxygen pair during and after each cascade, presented in Fig. \ref{fig:crystats}. The most significant change of the overall coordination distribution is related to the glass-forming Ti-O units, with a decrease in 6-coordinated ($^{[6]}$Ti) and an increase in 5- ($^{[5]}$Ti) and 7-coordinated ($^{[7]}$Ti) Ti atoms with O atoms. Ti-O demonstrates an initial quick recovery from the cascade spike in the number of $^{[6]}$Ti over the first few ps, but then a slower decrease in $^{[6]}$Ti to a new equilibrium structure over about 40 ps. Similar behaviour has been observed before in the Si-O glass forming unit in zircon  \cite{diver2020evolution}. DFT calculations \cite{mulroue2011ab} show that the most stable configuration for an oxygen interstitial in crystalline zirconolite creates 2 $^{[7]}$Ti ions. These interstitials introduce defect states into the band gap \cite{mulroue2011ab} which can be probed using optical techniques. It is also interesting to note Frenkel defect formation energies for O interstitial-vacancy pairs in zirconolite are relatively low, between 1.15 eV and 3.18 eV \cite{mulroue2011ab}.

Ca and Zr network modifiers show a much larger and faster recovery from the cascade induced spike, relaxing to a new equilibrium after approximately 20 ps. They also both show an overall decrease in the average coordination number, suggesting that displaced O atoms are forming vacancies on the network modifiers and interstitials on the Ti-O glass forming units.

The overall decrease in the coordination of Ca is not unexpected and has been seen before from x-ray absorption spectroscopy of fully damaged zirconolite \cite{lumpkin1986alpha}. It has been suggested that the coordination environment of Zr atoms is not significantly altered by radiation damage \cite{farges1993structure}, and while our results suggest that this is not the case, we do however observe that the magnitude of coordination change of Ca and Ti is significantly greater than is observed for Zr.

\subsection{Coordination statistics in amorphous zirconolite due to multiple overlapping cascades}

\begin{figure}
\centering
\includegraphics[width = 8cm]{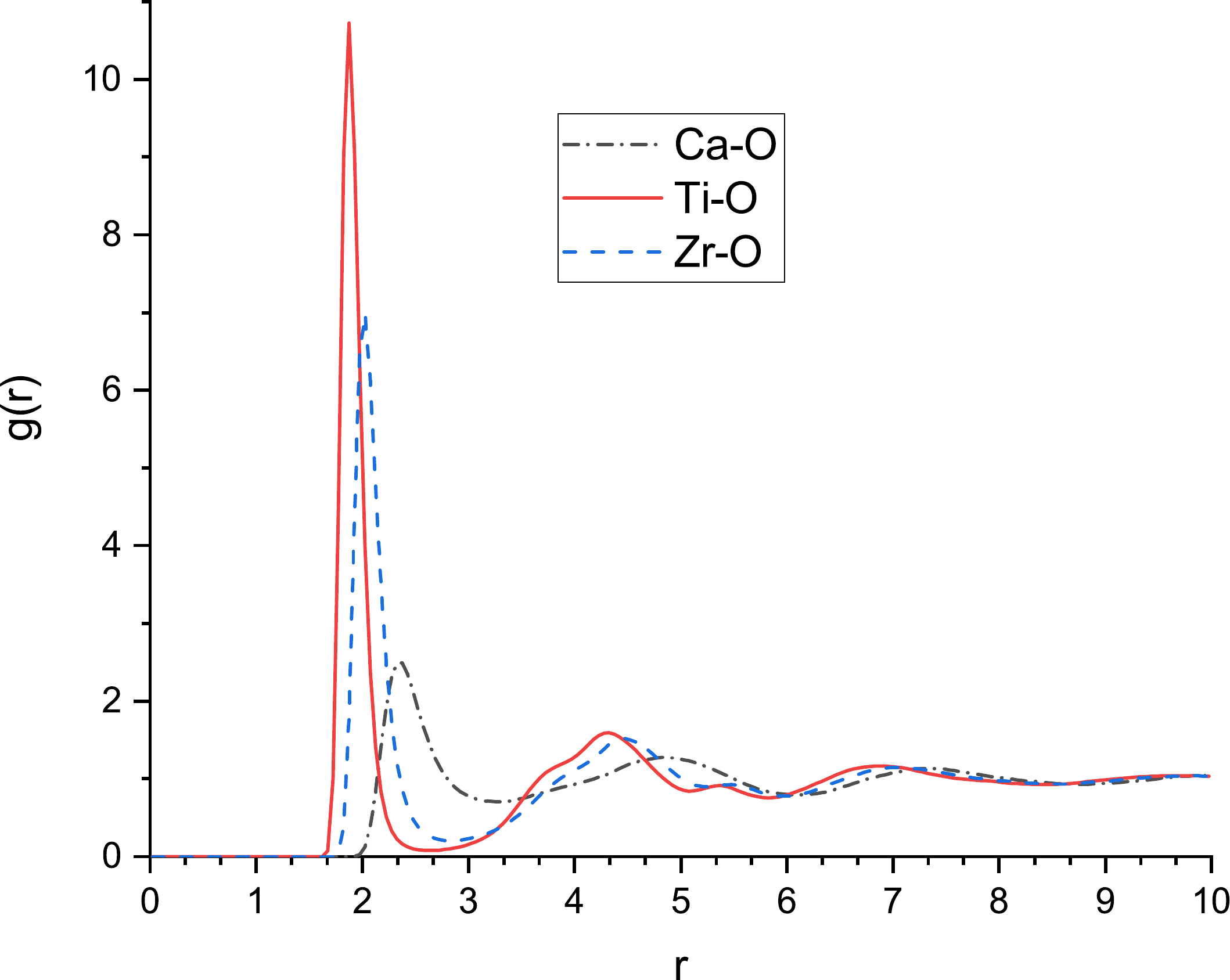}
\caption{Pair distribution functions of Ca-O, Zr-O and Ti-O pairs in amorphous zirconolite before any radiation damage}
\label{fig:grcoordpairsamr} 
\end{figure}

\begin{figure}
\centering
\includegraphics[width = 6cm]{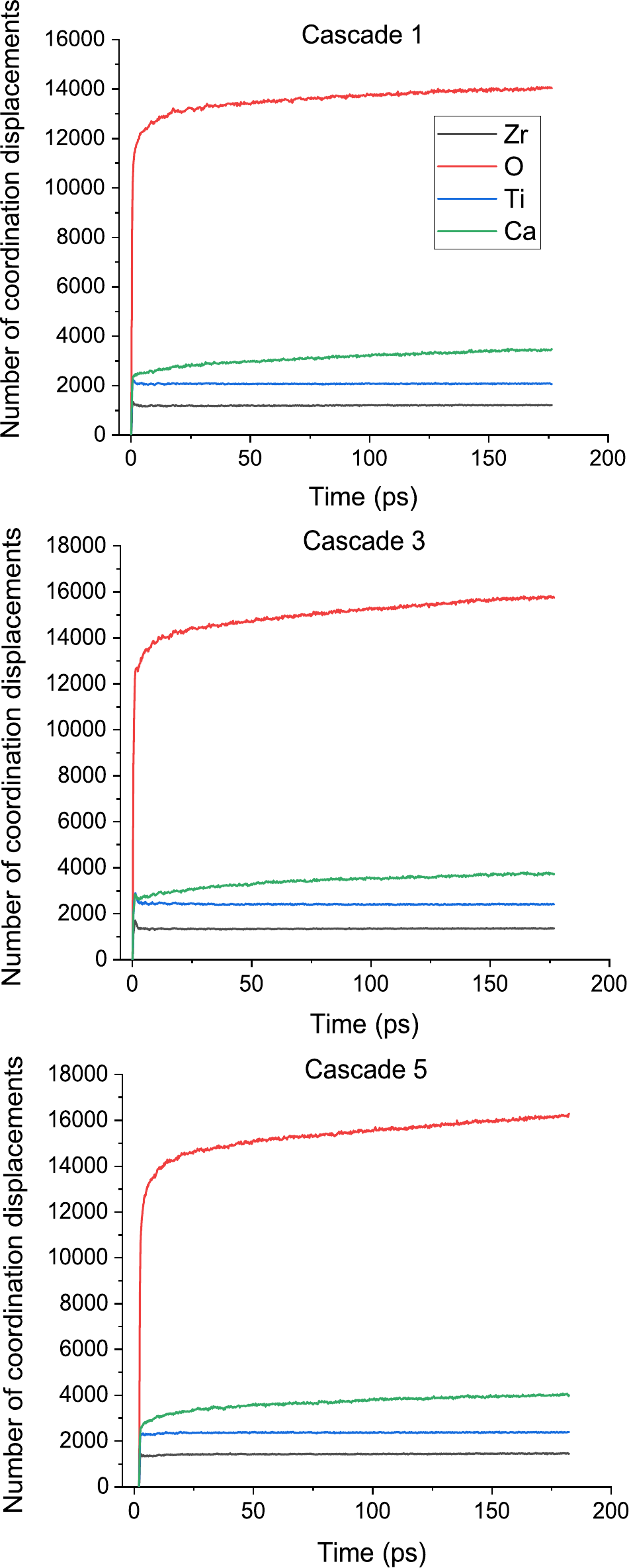}
\caption{Number of coordination displacements for Ti, O, Ca and Zr atoms during collision cascades 1,3 and 5 in crystalline zirconolite}
\label{fig:amrdis}
\end{figure}

Following on from the crystalline structure, we now analyze radiation cascades in amorphous zirconolite in much the same way as the crystal. We have determined different parameters to build local atomic coordinations, seen in Fig. \ref{fig:amrdis}, for the amorphous zirconolite structure. Cutoffs of 2.62 \si{\angstrom}, 2.87 \si{\angstrom} and 3.27 \si{\angstrom} are used for Ti-O, Zr-O and Ca-O pairs, respectively. These were determined from the first minimum after the first peak in the PDFs of the atom pairs as shown in Fig. \ref{fig:grcoordpairsamr}. The same values for the minimum displacement cutoffs in the amorphous material were selected as for the crystal to aid direct comparison.

The first point to note is the significantly increased number of coordination displacements that occur compared to the crystal. In both cases the more mobile oxygen atoms have a significantly larger number of displacements compared to the other atomic species, and overall 4 times more atoms are displaced during a cascade in amorphous zirconolite. In this sense, the amorphous structure becomes "softer" than the crystal in terms of higher susceptibility to displacements from radiation damage. We also observe that in the amorphous structure Ca atoms are the 2nd most highly displaced species after O, as opposed to the crystal where Ti atoms are the 2nd most displaced. This indicates that Ca ions are much more mobile in the amorphous structure, consistent with earlier results \cite{chappell2012structural}.

Another important feature of Fig. \ref{fig:amrdis} is the lack of any recovery peak in the first few picoseconds. This is in striking contrast to the behaviour seen in the crystal (Fig. \ref{fig:crydis}). This can be explained by the recovery being due to a reversible elastic deformation of the lattice surrounding the collision cascade core. Due to the strongly anharmonic motion produced inside the cascade core the region undergoes a volume expansion, which then causes tensile stress on the surrounding structure \cite{physrevbres}, resulting in only the temporary displacement of atoms on the outside of the cascade region. Our current results indicate that this elastic and reversible deformation operates at a much smaller scale in amorphous zirconolite. Once the damage reaches its peak, this peak determines all the subsequent damage in the structure. This is a notable result revealing an important difference in radiation response between crystalline and amorphous systems.

\begin{figure}[!htp]
\centering
\includegraphics[width = 10cm,keepaspectratio]{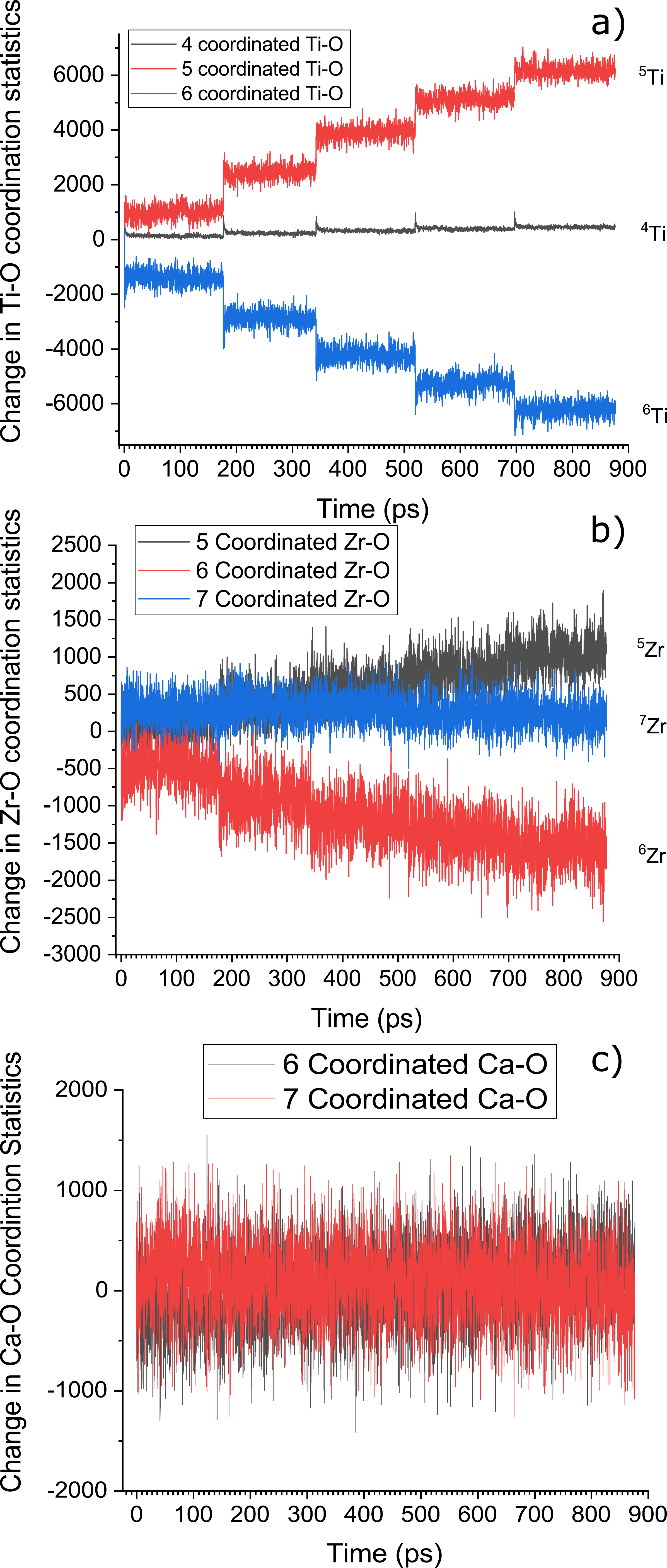}
\caption{Change in Zr-O coordination during consecutive collision cascades in amorphous zirconolite.}
\label{fig:amrstats}
\end{figure}

The coordination statistics for all pairs can be seen in Fig. \ref{fig:amrstats}, which show key features that are missed from only looking at displacements. From the Ca-O statistics we see no change in the overall coordination distribution due to radiation cascades. This suggests that the the Ca-O coordination distribution is in equilibrium, or 'saturated' in the amorphous state and that further damage does not continue to evolve the system to a new Ca-O coordination distribution. This process of saturation is also seen in some aspects of the Zr-O coordination statistics after the 2nd cascade. The number of newly created 7-coordinated Zr-O ($^{[7]}$Zr) stops increasing in the damaged region and remains roughly constant during the rest of the cascades. However, the number of $^{[5]}$Zr continues to increase. The divergence of the coordination statistics between $^{[5]}$Zr and $^{[7]}$Zr show evolution towards a new, stable amorphous structure produced by the collision cascades that will eventually reach a saturation limit where it can no longer be further damaged. This saturated state would be crucial to identify and study if zirconolite is going to be used for the long term storage of nuclear waste.

The amorphous structure produced via melt-quench has a lower initial average Ti-O coordination, 5.48, than the crystal. The Ti-O coordination distribution also undergoes a much larger change as a result of radiation damage, with a large increase in the number of $^{[5]}$Ti and a decrease in the number of $^{[6]}$Ti. Consistent with earlier findings, we see that it is easier to damage amorphous zirconolite and disrupt the Ti-O network than it is in the crystalline system. In the initial crystalline system the majority of the loss in the $^{[6]}$Ti atoms results in over coordinated $^{[7]}$Ti, but in the completely amorphous structure the $^{[6]}$Ti on average become $^{[5]}$Ti after subsequent cascades. This suggests that once complete amorphization of the crystal occurs, further cascades will still cause the structure to evolve towards a new "equilibrium" end state. At some point there is a switch when Ti coordination stops increasing and starts decreasing. It should be noted we do not observe the intermediate amorphous structure in the damaged crystalline zirconolite, even after 5 cascades, but this is likely due to the continuation of the structure evolution in the medium range (Fig. \ref{fig:crydamregiongr}) not reaching complete amorphization. Analysis of synthetic amorphized zirconolite produced from ion beam experiments \cite{reid2010structure} also shows an increase in the percentage of $^{[5]}$Ti. A similar result is reported in glass ceramic composites incorporating zirconolite where a decrease in Ti coordination number is observed as a result of structural damage \cite{paknahad2017investigation}. 

\subsection{Density inhomogeneity and energetic effects}

\begin{figure}
\centering
\includegraphics[width = 6cm]{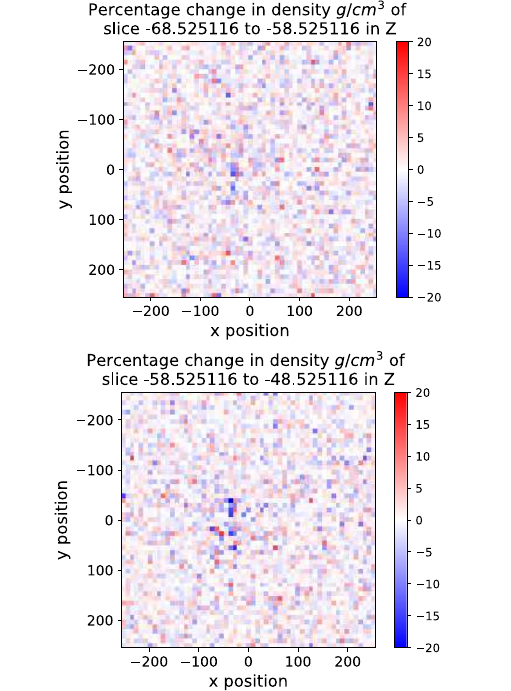}
\caption{Density fluctuation of amorphous zirconolite after the 1st collision cascade. The more red a region, the larger the increase in percentage change in density and the bluer a region the larger the decrease in percentage change in density}
\label{fig:amrbet1st}
\end{figure}

\begin{figure}
\centering
\includegraphics[width = 6cm]{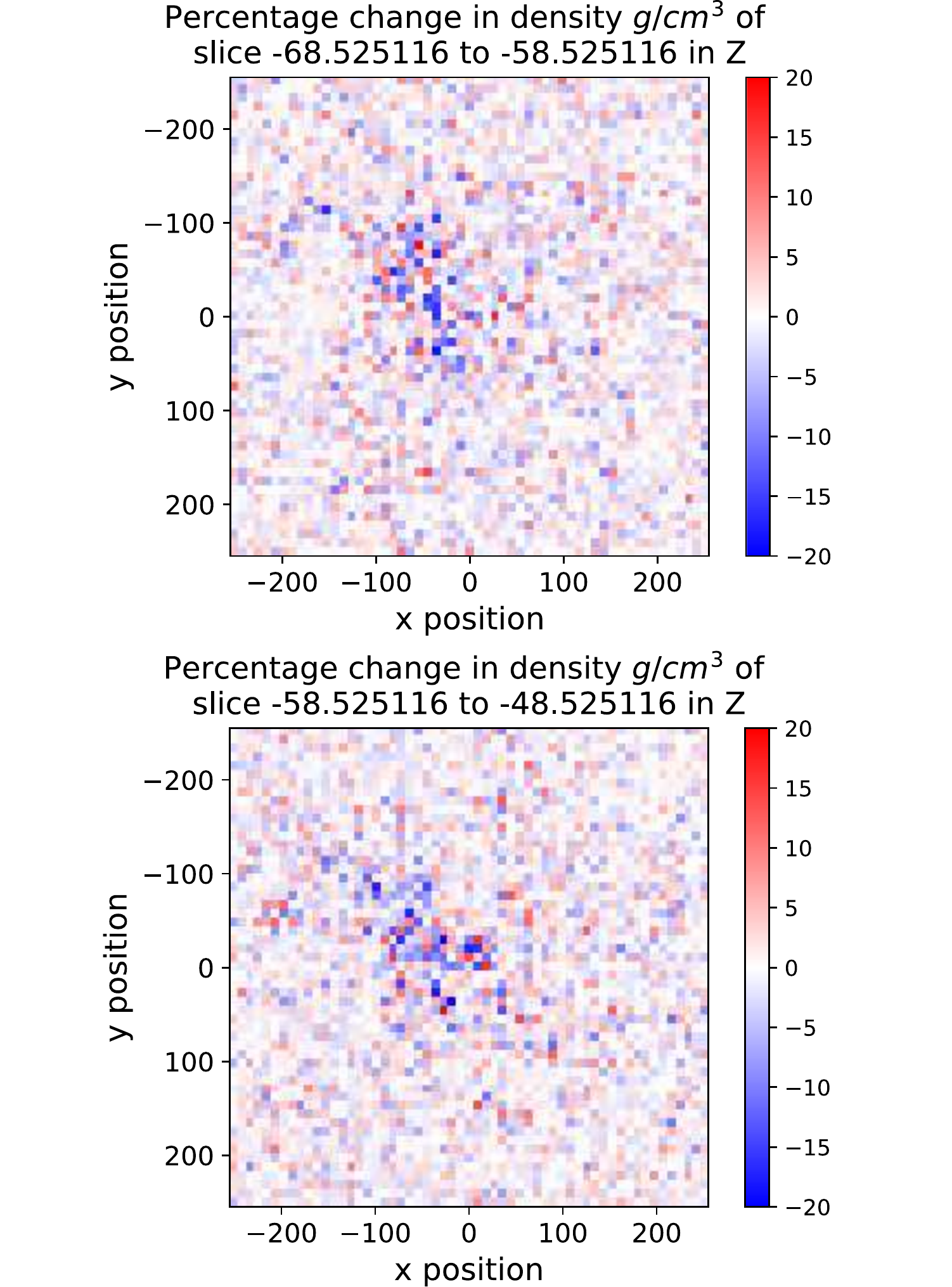}
\caption{Density fluctuation of amorphous zirconolite after the 1st collision cascade.The more red a region, the larger the increase in percentage change in density and the bluer a region the larger the decrease in percentage change in density}
\label{fig:amrbetall}
\end{figure}

In this section, we address radiation-induced density variations important for waste immobilization due to its potential effects on diffusion rates of radioactive ions. To measure this the simulation cell is split into boxes of \SI{10}{\angstrom} in length. The density of these boxes is calculated before each collision and then compared to the density after the cascade. As can be seen in Fig. \ref{fig:amrbet1st} and \ref{fig:amrbetall} we find an increase in density inhomegeneity in areas of the simulation box where the cascades are located. Although we see no discernible difference in the local PDFs of the damaged region, we clearly see a region with up to 20\% decreases in density, creating nanosized voids in the amorphous network as a result of radiation damage. Previous work examining natural radiation damaged zirconolites \cite{ewing1983alpha} also reported the existence of these voids, with sizes up to a hundred nanometers. The large void sizes measured experimentally may be related to the age of the samples and the formation of He bubbles as a result of the alpha decays, which further increase void size. Understanding these voids in terms of the performance of zirconolite as a waste form is paramount because they serve as fast diffusion pathways. This has been observed in Cm doped zirconolite where leaching rates up to 50 times larger than the crystalline structure have been measured after sample self irradiation \cite{weber1986effects}.

\begin{table}
\centering
\begin{tabular}{ccc}

{Change in enthalpy (eV)} &   Crystal  & Amorphous \\
\hline

Cascade 1& \si{722 \pm 34} & \si{1261 \pm 27} \\
\hline
Cascade 2 & \si{ 782 \pm 54 } &\si{1212 \pm 76} \\
\hline
Cascade 3 & \si{768 \pm 48} & \si{901 \pm 47} \\
\hline
Cascade 4 & \si{1190 \pm 59} & \si{926 \pm 56} \\
\hline
Cascade 5 & \si{1162 \pm 56} & \si{718 \pm 37} \\

\end{tabular}
\caption{Change in enthalpy between each consecutive collision cascade in amorphous and crystalline zirconolite}
\label{table:enthalpytest}

\end{table}

We calculate the enthalpy of the system progressively damaged by collision cascades. We first relax the initial crystalline and amorphous zirconolite structures with no radiation damage with an NPT ensemble for 50 ps and calculate average value for the total enthalpy of those structures. After each collision cascade we then also follow the same procedure. We then calculate a change in the enthalpy of the relaxed structures as a result of the radiation damage. In Table \ref{table:enthalpytest} we see that whilst total enthalpy increases in the crystal after each cascade, an opposite trend is seen the amorphous system. As zirconolite becomes amorphous the amount of enthalpy change increases with each cascade and as such it becomes easier for the structure to store energy in the above sense. On the other, it becomes increasingly more difficult to store the energy as the amorphous system is damaged further. This provides further support to the evolution of the crystalline phase towards an intermediate amorphous phase, where subsequent, continued damage will lead to the evolution of a different/new amorphous phase.


We observe that further cascades in the amorphous structure seem to have less impact on enthalpy change, suggesting that there may be a saturation in this sense. This is consistent with previous work in quartz highlighting the link between the decrease in density and the increase in enthalpy \cite{krishnan2017enthalpy} and suggesting that damage saturation is reached once the local enthalpy landscape is altered enough to produce lower energy barriers, such that further structurally induced defects are able to relax out.


From our results and analysis we propose that the change in enthalpy and coordination distribution of materials can be used to experimentally quantify radiation damage and allow comparison between materials.

\begin{figure}
\centering
\includegraphics[width = 6cm]{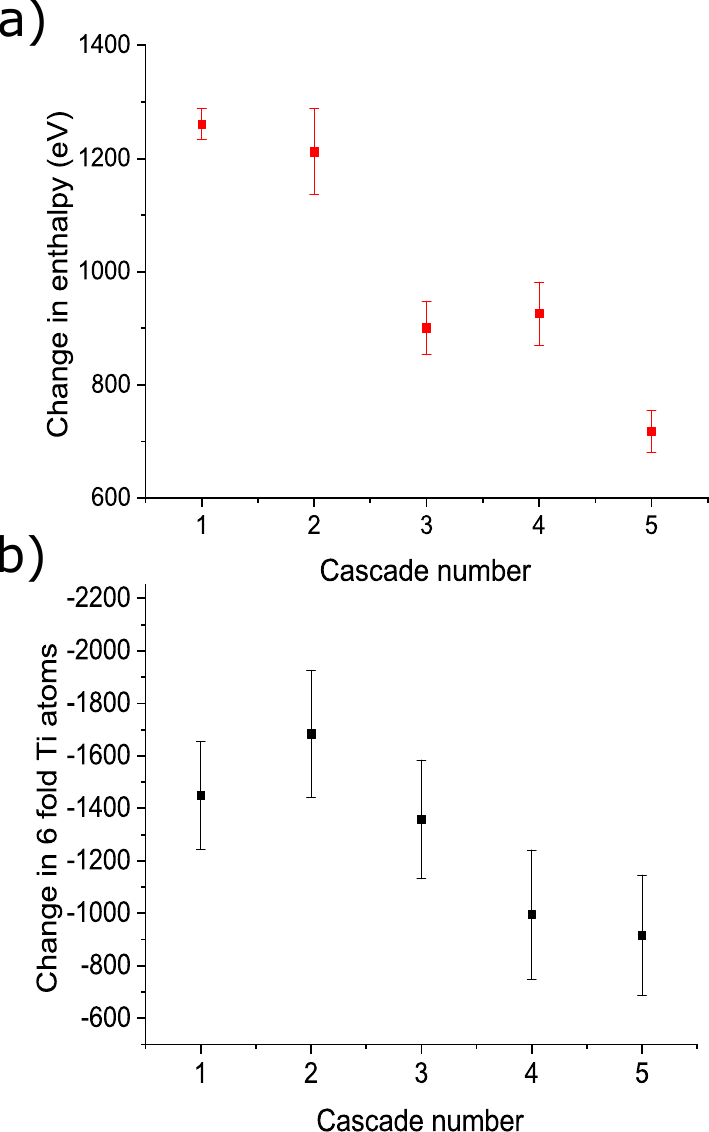}
\caption{Change in both a) system enthalpy and b) number of 6 coordinated Ti atoms between each collision cascade in amorphous zirconolite}
\label{fig:enthdam}
\end{figure}

A comparison of the change in enthalpy to the change in the number of $^{[6]}$Ti (Fig.\ref{fig:enthdam}) in amorphous zirconolite shows a sign of correlation which suggests that the change in enthalpy may be linked to the disruption of the Ti-O amorphous network. A similar correlation is observed in our recent work in zircon, with the disruption of the Si-O network \cite{diver2020evolution}. That this occurs in both titanates and silicates is an interesting result. It suggests the trend may be universal and not depend on the elemental composition of the glass.

\section{Conclusions}

In summary, we have studied radiation damage in amorphous zirconolite and found that the amorphous structure responds to radiation damage differently from the crystal: it is softer compared to the crystal in terms of defect recovery. We find that the amorphous structure continues to evolve as a result of radiation damage, whilst also identifying large density inhomogeneities at the local level which we suggest may play an important role for nuclear waste storage. Finally, we find a correlation between the changes of enthalpy and local coordination, suggesting that enthalpy change is a proxy for radiation damage. 

We also show that the evolution of coordination distribution act in opposite directions in the crystalline and amorphous phases with increasing damage, suggesting there is an intermediate amorphous phase that occurs once zirconolite is completely amorphized but not in long term equilibrium when subject to further cascade, Our results raise an interesting question of whether the detected evolution of the amorphous structure in response to radiation damage may {\it converge} to a new amorphous structure. This poses a question of fundamental importance to the physics of amorphous state \cite{zallenbook}, namely what is the nature of this new amorphous state? These questions can be studied in future experimental and modelling work. 

\section{Acknowledgements}

AD, OD and KT were supported by the UK Engineering and Physical Sciences Research Council (EPSRC) grant EP/R004870/1.

Via our membership of the UK's HEC Materials Chemistry Consortium, which is funded by EPSRC (EP/L000202, EP/R029431), this work used the ARCHER UK National Supercomputing Service (http://www.archer.ac.uk) and the UK Materials and Molecular Modelling Hub for computational resources, MMM Hub, which is partially funded by EPSRC (EP/P020194).

Part of this work made use of computational support by CoSeC, the Computational Science Centre for Research Communities, through CCP5: The Computer Simulation of Condensed Phases, EPSRC grant no EP/M022617/1 through AME and ITT.

\end{document}